\documentclass{iaus}
\usepackage{graphicx}
\usepackage{eufrak}

\title[Towards SSP models calibration] 
{Towards a calibration of SSP models from the optical to the mid-infrared}

\author[Pessev et al.]   
{P. Pessev$^1$, P. Goudfrooij$^1$, T. Puzia$^2$ and R. Chandar$^3$}

\affiliation{$^{1}$Space Telescope Science Institute, 3700 San Martin Drive, Baltimore, MD 21218, U.S.A. \\[\affilskip]$^{2}$Herzberg Institute of Astrophysics, 5071 West Saanich Road, Victoria, BC V9E 2E7, Canada\\[\affilskip]$^{3}$Carnegie Observatories, 813 Santa Barbara Street, Pasadena, CA 91101-1292, U.S.A. \\[3pt] emails: pessev@stsci.edu, goudfroo@stsci.edu, puziat@nrc.ca, rchandar@ociw.edu}

\pubyear{2007}
\volume{241}  
\pagerange{xxx--yyy}
\date{?? and in revised form ??}
\jname{Proceedings Stellar Populations as building blocks of galaxies}
\editors{A. Vazdekis \& R. Peletier, eds.}
\begin{document}

\maketitle

\begin{abstract}
Our knowledge about unresolved stellar systems comes from comparing integrated-light properties to SSP models. Therefore it is crucial to calibrate the latter as well as possible by integrated-light colors of clusters that have reliable ages and metallicities (deep CMDs and/or spectroscopy of individual giants). This is especially true for the NIR and MIR, which contain important population synthesis diagnostics and are often used to derive masses of stellar systems at hight redshifts. Here we present integrated colors of MC clusters using new $VRI$ photometry and 2MASS data. In the imminent future we will include MIR data from $Spitzer$/IRAC. We compare our new colors with popular SSP models to illustrate their strengths and weaknesses. 

\keywords{Magellanic Clouds, galaxies: star clusters, infrared: general, techniques: photometric}

\end{abstract}

{\bf Introduction} The concept of using populous star clusters as a testbed of Simple Stellar Population (SSP) models is not a new one (see \cite[Renzini \& Buzzoni 1986]{rb86}). Despite the gradual accumulation of observational data and knowledge about clusters, this information is still not fully utilized and dispersed in the literature. We present an ongoing project to provide a new multi-wavelength database of integrated-light properties of star clusters, covering a wide range in the age-metallicity parameter space. 

{\bf The dataset} consists of: {\it(i)} $UBV$ photoelectric data from the literature. {\it(ii)} New $VRI$ CCD imaging of 28 objects (\cite[Goudfrooij et al. 2006]{goudfrooij06}). {\it(iii)} 2MASS $JHK_S$ photometry of 75 clusters in the Magellanic Clouds (\cite[Pessev et al. 2006]{pessev06}). {\it(iv)} $Spitzer$/IRAC mid-infrared data for 6 of these clusters (Pessev et al. in preparation). The extended age/metallicity and wavelength coverage makes these objects ideal to test the existing and calibrate  the upcoming SSP models.  As an illustration of the data, we combine optical and NIR photometry of old MC clusters to test recent, widely used SSP models (\cite[Maraston 2005]{ma05}, \cite[Bruzual \& Charlot 2003]{bc03}, \cite[Anders \& Fritze 2003]{af03}, \cite[Vazdekis 2000]{va99}).

\begin{figure}
\begin{center}
\includegraphics[bb=0 127 595 693,width=11.8cm]{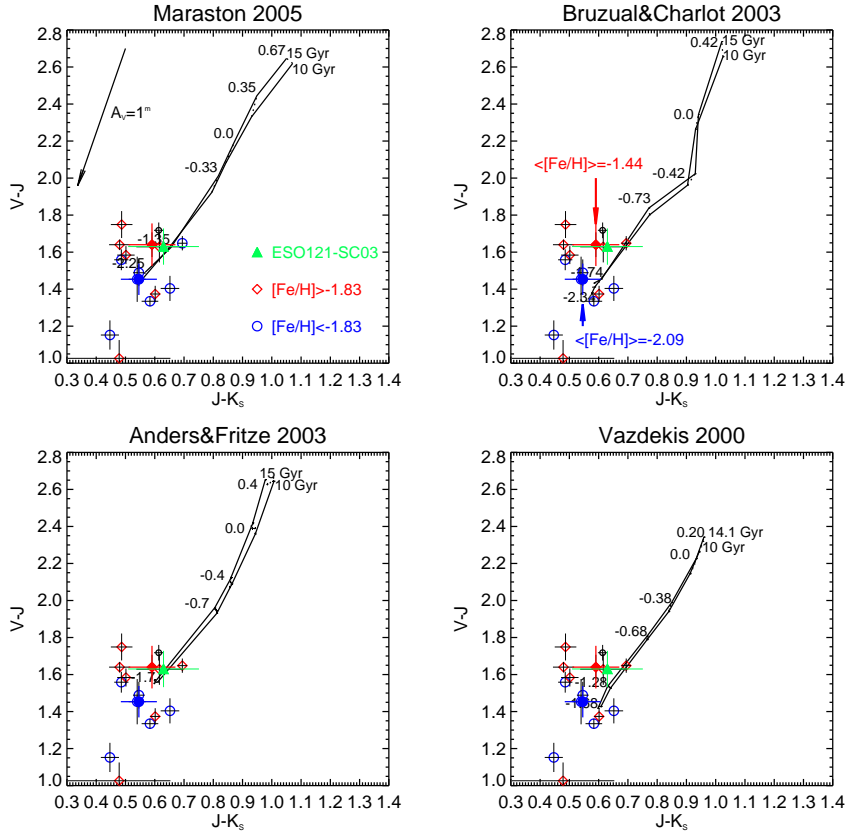}
\caption{The model grids for two different ages (metallicity and age are shown along the model sequences) and our data points (filled symbols represent the binned data). }
\label{fig:one}
\end{center}
\end{figure}

{\bf Testing the models} The model colors were transformed to the 2MASS system and $(V-J)$ vs. $(J-K_S)$ diagrams covering appropriate parameter space are presented on Figure~\ref{fig:one}. The color combination provides high efficiency, metallicity resolution, and is close to the wavelength coverage of the upcoming telescopes. Photometry information for 15 old MC clusters (Age $\ge$ 10 Gyr) is overlaid on the model grids\footnote{The Reticulum cluster is not considered due to insufficient depth of 2MASS imaging data. }, along with ESO121-SC03 (the only cluster in the LMC "age gap"). Old MC clusters are massive enough (\cite[McLaughlin \& van der Marel 2005]{mclvdm05}) to minimize the intrinsic spread along the model tracks due to stochastic fluctuations in the stellar populations ($log(\mathfrak{M/M}_{\bigodot})\ge5$; \cite[Bruzual 2002]{bruzual02}). To further increase accuracy, we divide the sample in metallicity, accumulating more than $10^{6}\mathfrak{M}_{\bigodot}$ in each bin and derive mean parameters of the subsamples.

{\bf Conclusions} We find a good agreement between the predictions of all models and the observations. \cite[Maraston 2005]{ma05} models with blue horizontal branches in particular, seem to better reproduce old metal-poor MC clusters. The combination of optical/NIR colors can be used to detect and analyze old  GC populations distant galaxies.

\begin{acknowledgments}
Support for this work was provided in part by NASA through a Spitzer Space Telescope Program, through a contract issued by the JPL, CIT under a contract with NASA. 

P.P. is thankful for the financial support by IAU through the LOC of IAUS 241.
\end{acknowledgments}


\begin{thebibliography}{}
\bibitem[Anders \& Fritze (2003)]{af03} 
              {Anders, P., \& Fritze, U.\ } 2003, 
              \textit{A\&A}, 401, 1063
\bibitem[Bruzual A.(2002)]{bruzual02} 
	     {Bruzual A., G.\ }2002, 
	     \textit{IAU Symposium, 207}, 616
\bibitem[Bruzual \& Charlot(2003)]{bc03} 
	     {Bruzual, G., \& Charlot, S.\ } 2003, 
	     \textit{MNRAS}, 344, 1000
\bibitem[Goudfrooij et al.(2006)]{goudfrooij06} 
              {Goudfrooij, P., Gilmore, D., Kissler-Patig, M., \& Maraston, C.\ } 2006,
              \textit {MNRAS}, 369, 697 
\bibitem[McLaughlin \& van der Marel(2005)]{mclvdm05} 
              {McLaughlin, D.~E., \& van der Marel, R.~P.\ } 2005, 
              \textit{ApJS}, 161, 304 
\bibitem[Maraston(2005)]{ma05} 
	     {Maraston, C.\ } 2005, 
	     \textit{MNRAS}, 362, 799
\bibitem[Pessev et al.(2006)]{pessev06} 
	     {Pessev, P.~M., Goudfrooij, P., Puzia, T.~H., \& Chandar, R.\  }2006, 
	     \textit{AJ}, 132, 781
\bibitem[Renzini \& Buzzoni(1986)]{rb86} 
	    {Renzini, A., \& Buzzoni, A.\ }1986, 
	    \textit{ASSL Vol.~122: Spectral Evolution of Galaxies}, 195
\bibitem[Vazdekis(1999)]{va99} 
	     {Vazdekis, A.\ }1999, 
	     \textit{ApJ}, 513, 224	
	    \end{thebibliography}
\end{document}